\documentclass[superscriptaddress,aps,pra,twocolumn,showpacs,preprintnumbers,amsmath,amssymb,floatfix]{revtex4-1}
\usepackage{graphicx}
\usepackage{dcolumn}
\usepackage{bm}
\usepackage{color}

\usepackage{braket}

\def\d{\downarrow}
\def\u{\uparrow}

\def\u{\uparrow}
\def\d{\downarrow}

\def\ba{\begin{eqnarray}}
\def\ea{\end{eqnarray}}
\def\beq{\begin{equation}}
\def\eeq{\end{equation}}

\def\be{\begin{equation}}
\def\ee{\end{equation}}
\def\bea{\begin{eqnarray}}
\def\eea{\end{eqnarray}}

\usepackage{subfiles}

\begin{document}

\title{Topological phases in ultracold polar-molecule quantum magnets}


\author{Salvatore R. Manmana}
\affiliation{Institute for Theoretical Physics, University of G\"ottingen, D-37077 G\"ottingen, Germany}
\affiliation{JILA, NIST and Department of Physics, University of Colorado, Boulder, CO 80309, USA}
\affiliation{Kavli Institute for Theoretical Physics, University of California, Santa Barbara, CA 93106, USA}
\author{E. M. Stoudenmire}
\affiliation{Department of Physics and Astronomy, University of California, Irvine, CA 92697, USA}
\author{Kaden R. A. Hazzard}
\affiliation{JILA, NIST and Department of Physics, University of Colorado, Boulder, CO 80309, USA}
\affiliation{Kavli Institute for Theoretical Physics, University of California, Santa Barbara, CA 93106, USA}
\author{Ana Maria Rey}
\affiliation{JILA, NIST and Department of Physics, University of Colorado, Boulder, CO 80309, USA}
\author{Alexey V. Gorshkov}
\affiliation{Institute for Quantum Information \& Matter, California Institute of Technology, Pasadena, CA 91125, USA}

\date{\today}

\begin{abstract}

We show how to use polar molecules in an optical lattice to engineer quantum spin models with arbitrary spin $S\ge 1/2$ and with interactions featuring a direction-dependent spin anisotropy. 
This is achieved by encoding the effective spin degrees of freedom  in microwave-dressed rotational states of the molecules and by coupling the spins through dipolar interactions. 
We demonstrate how one of the experimentally most accessible anisotropies stabilizes symmetry protected topological phases in spin ladders.
Using the numerically exact density matrix renormalization group method, we find  that these interacting phases -- previously studied only in the nearest-neighbor case -- survive in the presence of long-range dipolar interactions.
We also show how to use our approach to realize the bilinear-biquadratic spin-1 
and the Kitaev honeycomb models. 
Experimental detection schemes and imperfections are discussed.

\end{abstract}


\maketitle


Recent advances in ultracold polar molecules \cite{ni08, aikawa10,deiglmayr08}, Rydberg atoms \cite{saffman10,schauss12}, magnetic atoms \cite{aikawa12,lu12}, and magnetic defects in solids \cite{childress06,balasubramanian09,weber10} have spurred tremendous
interest in
exotic strongly-correlated many-body phenomena  
arising from anisotropic, long-ranged dipole-dipole interactions \cite{baranov08,lahaye09,trefzger11,baranov12,barnett06,micheli06, brennen07,buchler07b,gorshkov08c,wall09,wall10,schachenmayer10,perezrios10,herrera10,weimer10, dalmonte10,pohl10,ospelkaus10,demiranda10,kestner11,lemeshko11,gorshkov11b,gorshkov11c,mathey11,zhou11,dalmonte11b,babadi11,chotia12,weimer12,hazzard12b,lemeshko12,sowinski12,maik12}. 
The types of 
anisotropies realizable with these interactions are typically limited to simple changes of the interaction sign and magnitude according to the 
spherical harmonic $Y_{2,0} \propto 1 - 3 \cos^2 \theta$, where
$(\theta,\phi)$ are the spherical coordinates of the vector connecting the two interacting dipoles \cite{gorshkov11b,gorshkov11c,baranov08,trefzger11}.

In this Letter we show, in the context of polar molecules, that microwave dressing provides a tremendous degree of simultaneous control 
over five independent dipole-dipole interaction terms whose angular 
dependences are given by the 
rank-2 spherical harmonics.
This opens the door to simulating well-known models including the spin-1/2 XXZ model 
with a direction-dependent spin anisotropy, the spin-1 bilinear-biquadratic model \cite{schollwock96}, and the Kitaev honeycomb model \cite{kitaev06}. Thanks to the use of direct dipole-dipole coupling, the resulting interactions 
are stronger and hence easier to observe experimentally than other -- potentially direction-dependent -- spin-spin interactions such as  superexchange in ultracold atoms \cite{bloch08} or perturbative dipole-dipole-mediated
couplings between polar molecules \cite{micheli06,brennen07}.

As a specific example 
demonstrating the reach of our method,  we show how to design a spin-1/2 XXZ model with direction-dependent 
spin anisotropy using a minimal and experimentally reasonable microwave configuration. 
In a two-legged ladder geometry with nearest-neighbor interactions, this model has been shown to exhibit  symmetry protected topological (SPT) phases \cite{liu12b}. 
These phases are exotic gapped states of matter distinct from trivial gapped phases when specific symmetries are present.
They have recently attracted extensive interest \cite{schnyder08,schnyder09,kitaev09,ryu10,gu09,chen11e,chen12b,pollmann12b,schuch11b}  because
they do not fit within the framework of Landau symmetry breaking and possess exotic properties such as topologically protected edge states \cite{alicea12}, nonlocal order parameters \cite{dallatorre06,endres11}, and unique entanglement properties \cite{pollmann10,chen12b}. Using the density matrix renormalization group method (DMRG) \cite{schollwock05}, we compute the phase diagram of the two-legged-ladder model obtained in our polar molecule implementation and provide evidence that --  at least in this one-dimensional model -- SPT phases also exist in the presence of long-range dipolar interactions.

In a major advance over Refs.~\cite{gorshkov11b,gorshkov11c,yao12d}, our proposal realizes an interacting topological phase.  Furthermore, relying 
on 
homogeneous microwave -- not optical -- dressing, our proposal is much easier to realize experimentally  than the one on topological flatbands \cite{yao12d}. 
Finally, since the relevant motional energy scale in our setup is the lattice bandgap, our proposal can 
be realized at much higher motional temperature than the $t$-$J$-type model of Refs.~\cite{gorshkov11b,gorshkov11c}, which relies on tunneling. 



\begin{figure}[t]
\begin{center}
\includegraphics[width = 0.7 \columnwidth]{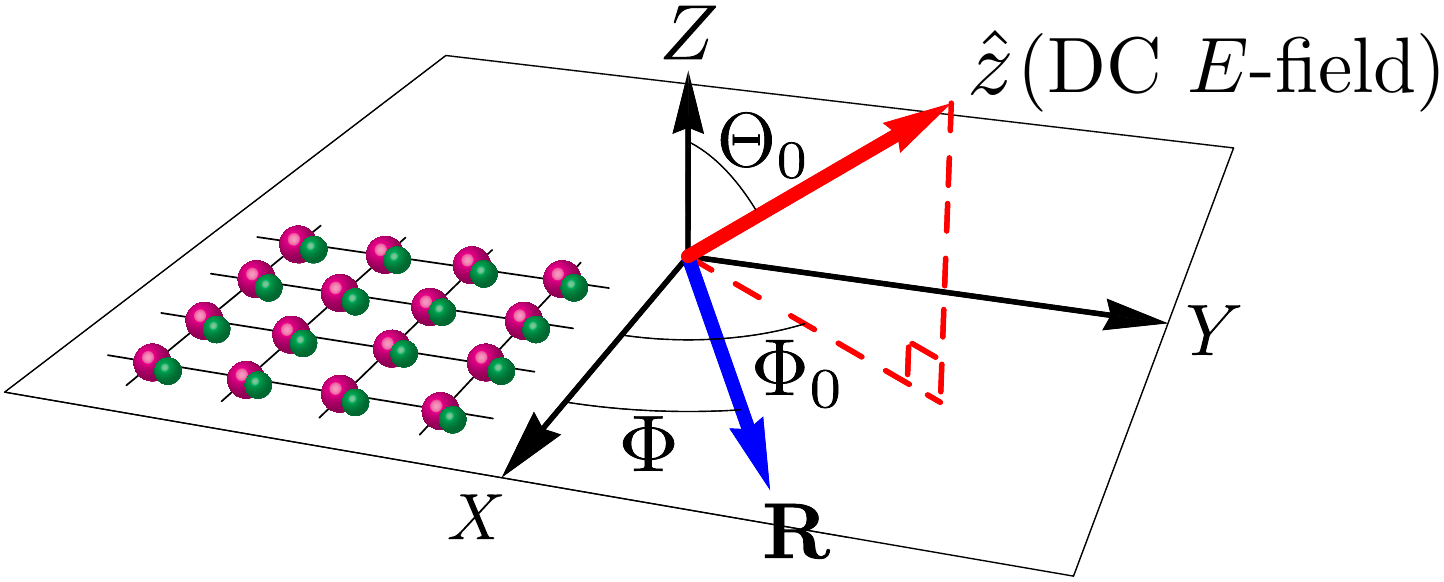}
\caption{(color online). 
A lattice of polar molecules in the $XY$ plane 
is subjected to a DC electric field along $\hat z$. We define the $xyz$ coordinate system as the rotation of the $XYZ$ coordinate system around $\hat Z$ by $\Phi_0$ and then around $\hat y$ by $\Theta_0$. A  vector $\mathbf{R}$ with polar coordinates $(R,\Phi)$ in the $XY$ plane has spherical coordinates  $(R,\theta, \phi)$ in the $xyz$ coordinate system. 
 \label{fig:scheme}}
\end{center}
\end{figure}

\textit{Setup.}---
 We consider an array of polar molecules confined  to the $XY$ plane and pinned in a deep optical lattice with one molecule per site [see Fig.~\ref{fig:scheme}(a)].
 Each molecule is treated as a rigid rotor with dipole moment operator $\mathbf{d}$, angular momentum operator $\mathbf{N}$, and a rotational constant $B$ and is described by 
the Hamiltonian $H_0 = B \mathbf{N}^2-E d^z$ in the presence of a DC electric field $E$ along $\hat z$.  
As one turns on  $E$, the simultaneous eigenstates of $\mathbf{N}^2$ and $N_z$ with eigenvalues $N (N+1)$ and $M$ adiabatically connect to eigenstates of 
$H_0$, which we denote $|N,M\rangle$. 
The dipole-dipole interaction between molecules $i$ and $j$ separated by $\mathbf{R}$ [see Fig.~\ref{fig:scheme}(a)] 
 is \cite{brown03}
\ba
H_{ij} &=& - \frac{\sqrt{6}}{R^3} \sum_{q = -2}^{2} (-1)^q C_{-q}^2(\theta, \phi) T^2_q(\mathbf{d}_i, \mathbf{d}_j), \label{eq:dd}
\ea
where \mbox{$C^2_q(\theta, \phi)= \sqrt{4 \pi/5}\ Y_{2,q}(\theta,\phi)$}
and the many-body Hamiltonian is $H=(1/2) \sum_{i\ne j} H_{ij}$. 
Here $T^2_q(\mathbf{d}_i, \mathbf{d}_j)$ is given by
\mbox{$T^2_{\pm 2}  = d_i^\pm  d_j^\pm$},
\mbox{$T^2_{\pm 1}  = \left(d_i^0  d_j^\pm + d_i^\pm  d_j^0\right)/\sqrt{2}$}, and
\mbox{$T^2_0= \left(d_i^-  d_j^+ + 2 d_i^0  d_j^0 + d_i^+  d_j^-\right)/\sqrt{6}$},
 where $d^0 = d^z$
and  \mbox{$d^\pm =  \mp (d^x \pm i d^y)/\sqrt{2}$}.
Thus $T^2_q$ changes the total $M$ of the two molecules by $q$. For $R \sim 0.4$ $\mu$m, the interaction energy scale is $d^2/R^3 \sim 1\;(100)$ kHz in KRb (LiCs).

To obtain a spin-$S$ Hamiltonian, we select in each molecule $2 S + 1$ disjoint sets of $|N,M\rangle$ states and couple the states within each set 
  to form dressed states  (\mbox{$\sim n$} microwave fields are needed to couple $n$ states). We then choose one 
  \cite{Note1}  
  dressed state from each set to create the spin-$S$ configuration.
Projecting Eq.~(\ref{eq:dd}) onto the chosen spin-$S$ basis, the resulting spin-spin interactions consist of
five potentially independently controllable terms with angular 
dependences 
 $C^2_0$, $\textrm{Re}[C^2_1]$, $\textrm{Im}[C^2_1]$, $\textrm{Re}[C^2_2]$, and $\textrm{Im}[C^2_2]$.
Refs.\ \cite{gorshkov11b,gorshkov11c} considered  
the special case of $S=1/2$ in the presence of the $C^2_0$ term alone and limited the discussion to situations where total $S^z$ was conserved. 
In this Letter, we evince the power of the approach beyond this special case. 




\textit{Interactions featuring a direction-dependent spin anisotropy.}---Our first demonstration of novel direction-dependent interactions focuses  on $S = 1/2$ and 
assumes that $H_{ij}$ connects a pair of molecules in the state $|m_1\rangle |m_2\rangle$ ($\ket{m_1}$ and $\ket{m_2}$ are dressed states)  only to itself and to $|m_2\rangle| m_1\rangle$, while all the other processes are off-resonant and thus negligible.
Although one may be able to independently control each of the five $C^2_q$ terms, 
here we will focus  on the  $C^2_{0,\pm 2}$ terms since $C^2_{\pm 1}$ terms are resonant only at specific values of $E$ 
\cite{Note2}. 

Consider the 
level configuration in Fig.~\ref{fig:levs}(a). 
We assume \mbox{$\Omega_\pm > 0$} and \mbox{$|\Omega_\pm| \gg H_{ij}$} and  take 
\mbox{$\ket{\u} = \ket{0,0}$} and $\ket{\d}  =\alpha |1,-1\rangle - \beta |1,1\rangle$ as our 
dressed spin states, where $\left\{\alpha, \beta\right\} = \left\{\Omega_-,\Omega_+\right\}/\sqrt{\Omega_-^2 + \Omega_+^2}$.
 Notice that $\ket{\d}$ is a single-molecule eigenstate in the presence of $\Omega_\pm$.
The spin model is then derived by projecting $H_{ij}$ onto states $\ket{\u}$ and $\ket{\d}$ via the same steps as in Ref.\ \cite{gorshkov11c} with one major difference:  \mbox{$d^+_i d^+_j$}, featuring a $C^2_{-2}$ angular dependence, resonantly couples \mbox{$|1,-1\rangle |0,0\rangle \rightarrow |0,0\rangle |1,1\rangle$}. The result is \cite{supp}
\ba
R^3 H_{ij} = J_z(\Phi)  S^z_{i} S^z_{j} + J_{xy}(\Phi) (S^x_i S^x_j  + S^y_i S^y_j),\label{eq:XXZ}
\ea
where $J_z(\Phi) = (1 - 3 \cos^2 (\Phi-\Phi_0) \sin^2 \Theta_0) (\mu_0 - \mu_1)^2$, $J_{xy}(\Phi) = - \mu_{01}^2 (1 - 3 \cos^2 (\Phi-\Phi_0) \sin^2 \Theta_0) +6 \alpha \beta \mu_{01}^2 [1 - \cos^2 (\Phi-\Phi_0) (1 +\cos^2 \Theta_0)]$, $\mu_0 = \langle 0,0|d^0|0,0\rangle$, $\mu_1 = \langle 1,1|d^0|1,1\rangle$,  $\mu_{01} =  \langle1,1|d^+|0,0\rangle$, and $\mathbf{S}_j$ is the 
spin-1/2 operator for molecule $j$.
The spin anisotropy $J_{xy}/J_z$ of this XXZ model changes depending on the polar angle $\Phi$ of the vector $\mathbf{R}$ connecting the two interacting molecules. 
As we discuss below, Eq.~(\ref{eq:XXZ}) allows one to study SPT phases in ladders.  Another 
special case is a 
square-lattice Heisenberg model 
with a tunable ratio between coupling strengths on $\hat X$ and $\hat Y$ bonds~\cite{kim:monte_2000}.
In the nearest-neighbor limit, 
this 
enables one to study 
the change from one-dimensional (uncoupled) chains to a two-dimensional behavior.  Such models have also been used to explore the physics of stripes in high-temperature superconductors~\cite{neto:doped_1996,vanDuin:charge_1998}. 
While we see that even the simple level structure of Fig.~\ref{fig:levs}(a) yields a wealth of exotic physics, we show in the next two sections that additional features can be accessed with increased microwave control.

\begin{figure}[t]
\begin{center}
\includegraphics[width = 0.87 \columnwidth]{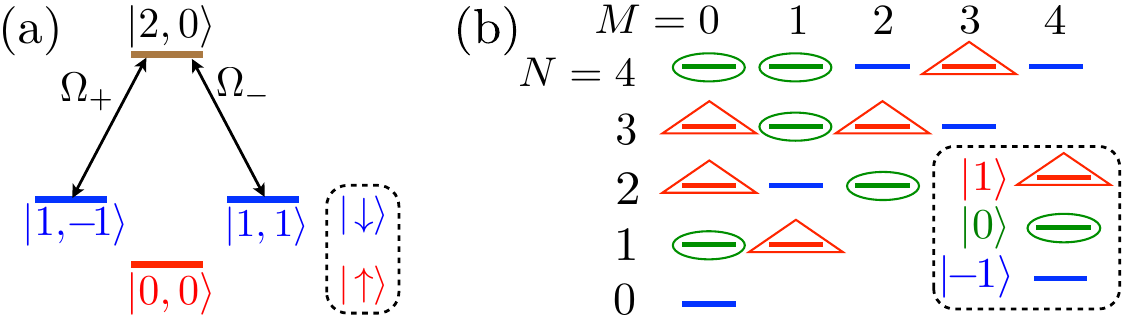}
\caption{(color online).  (a) The level scheme and resonant microwave coupling used to realize the Hamiltonian, Eq.~(\ref{eq:XXZ}). 
The dressed states we choose are $\{\ket{\u},\ket{\d}\} = \{\ket{0,0}, (\Omega_- \ket{1,\!-\!1} - \Omega_+ \ket{1,1})/\sqrt{\Omega_-^2 + \Omega_+^2}\}$. 
(b) Microwave-dressed rotational levels for the SU(2)-symmetric spin-1 model.  The dressed states are $|1\rangle$ (linear combination of states indicated by triangles), $|0\rangle$ (ovals), and $|\!-\!1\rangle$ (the rest).  The diagram is schematic: the real system is anharmonic and levels $|N,M\rangle$ with the same $N$ are non-degenerate (unless the levels have the same $|M|$).  
  \label{fig:levs}}
\end{center}
\end{figure}

\textit{Degenerate dressed states and non-Abelian anyons.}--- To realize models such as the quantum compass model \cite{kugel73}, the Kitaev honeycomb model \cite{kitaev06}, and the Yao-Kivelson model \cite{yao07}, we need to go beyond Eq.~(\ref{eq:XXZ}) and realize terms, such as $S^x_i S^x_j$, that do not conserve the total $S^z$. 
To do this, we simply tune $\ket{\u}$ and $\ket{\d}$ to be degenerate. 

As an example, consider the Kitaev honeycomb model, where interactions along \mbox{$\Phi = \pi/6, \pi/2$}, and $5 \pi/6$ are of the form $S^x_i S^x_j$, $S^y_i S^y_j$, and $S^z_i S^z_j$, respectively \cite{kitaev06}. At $(\Theta_0,\Phi_0) = (0,0)$, 
the interaction between two molecules $i$ and $j$ is 
\mbox{$R^3 H_{ij}(\Phi) = \mathbf{v}(\Phi) \cdot \mathbf{M}$}, 
where \mbox{$\mathbf{v}(\Phi) = \left\{1,-3 \cos(2 \Phi)/2, 3 \sin(2 \Phi)/2\right\}$} 
and \mbox{$\mathbf{M} = \{\sqrt{3/2} T^2_0, T^2_2 + T^2_{-2}, i T^2_2 - i T^2_{-2}\}$}. Since $\mathbf{v}(\pi/6)$, $\mathbf{v}(\pi/2)$, and $\mathbf{v}(5 \pi/6)$ are linearly independent, it is, in principle, possible to choose
the degenerate dressed states $\ket{\u}$ and $\ket{\d}$ to ensure that \mbox{$H_{ij}(\pi/6) \propto S^x_i S^x_j$}, \mbox{$H_{ij}(\pi/2) \propto S^y_i S^y_j$}, and 
\mbox{$H_{ij}(5 \pi/6) \propto S^z_i S^z_j$}. 
In Ref.~\cite{inprep}, we show that -- with sufficient microwave control -- such a choice of  dressed states is indeed possible, allowing one to realize 
the Kitaev B phase in the presence of a magnetic field. This gapped phase  supports non-Abelian anyonic excitations, which can be used, for example, for topologically protected quantum state transfer \cite{yao11c} and quantum computing \cite{kitaev06}. 

\textit{$S > 1/2$ and the bilinear-biquadratic model.}---We now show that
one can extend this tremendous control over spin-spin interactions
to \mbox{$S>1/2$}. In particular, we show how to obtain the general SU($2$)-symmetric spin-1 Hamiltonian, i.e.\ the bilinear-biquadratic Hamiltonian $\cos (\gamma) \mathbf{S}_i \cdot \mathbf{S}_j + \sin(\gamma) (\mathbf{S}_i \cdot \mathbf{S}_j)^2$,
which has a rich phase diagram even in one dimension  \cite{schollwock96,garcia-ripoll04,brennen07}. 
In particular, $\gamma = \pi/4$ and $\arctan(1/3)$ give the SU($3$)-symmetric and the AKLT  \cite{affleck87} Hamiltonians, respectively. One can also consider generalizations to SU($N$) with arbitrary $N$ as well as away from SU($2$) \cite{berg08}.

We build our spin-1 dressed-state basis $\{\ket{1},\ket{0},\ket{-1}\}$ from the 15 bare levels shown in Fig.~\ref{fig:levs}(b). For simplicity, the model presented here will have only the $C^2_0$ term. We work at 
$E = 3.244 B/d$ [$=13 \;(7)$ kV/cm  in KRb (LiCs)], 
for which 
the bare-states process  $\ket{1,0}\ket{1,0} \!\! \rightarrow \!\! \ket{2,0} \ket{0,0}$ is resonant. Furthermore, we choose the energies of the dressed states to make the process $\ket{0}\ket{0} \! \rightarrow \ket{-1}\ket{1}$ resonant \cite{Note3}. 
 The latter is needed to engineer the  
$S^-_iS^+_j$ term present in the desired Hamiltonian.
Aside from this exception, we again assume that a pair of molecules in dressed states $|m_1\rangle |m_2\rangle$ is connected via $H_{ij}$ only to itself and to $|m_2\rangle |m_1\rangle$.
Using 12 microwaves to create the dressed states in Fig.~\ref{fig:levs}(b),
we find \cite{supp} that we can achieve any $\gamma$ and, thus,  any bilinear-biquadratic Hamiltonian. Removing all five $N = 4$ 
states, we are left with just 8 microwaves, which simplifies the experimental  implementation but at the cost of only accessing the  $\gamma = 1.1$ point. 
The typical strength of interactions achieved is $R^3 H_{ij} \sim 0.01 d^2$. Stronger interactions and a reduced number of 
microwaves might be achievable by further optimizing the choice of levels and  microwaves.


\textit{SPT phases in spin ladders.}---
We now turn to the focus of our work: we use Eq.~(\ref{eq:XXZ}) to implement a specific  ladder model [see Fig.~\ref{fig:SPT}(a)] introduced in Ref.\  \cite{liu12b}   and shown to support nontrivial SPT phases for  nearest-neighbor interactions.  The symmetries protecting these interacting topological phases 
are the exchange $\sigma$ of the two legs and $D_2 = \{E,R_x,R_y,R_z\}$, where $E$ is the identity and $R_\alpha$ is a $\pi$-pulse around the axis $\alpha$ on all spins.
 Note that  although we focus 
on a ladder system since it is amenable to a numerically exact treatment, we expect an even  richer phase diagram in dimensions $D > 1$.



 While for our choice of levels $b \equiv \mu_{01}^2/(\mu_0 - \mu_1)^2$ satisfies $b \in [2.6, \infty)$, 
any $b \geq 0$ can be accessed by using reduced nuclear spin overlaps between $\ket{\u}$ and $\ket{\d}$ (see ``Experimental considerations" below) or by choosing $\ket{N,M}$ with different $N$.
To ensure the $\sigma$ symmetry, 
we take $\Phi_0 = \pi/2$.
To ensure a Heisenberg Hamiltonian 
along $\Phi = 0$,
we choose $\alpha \beta = (b + 1)/(6 b)$.
Since $\alpha \beta \in [0,1/2]$ and  $b \geq 0$,  
$b$ can go from $1/2$ to $\infty$.  Rescaling the interaction by $(\mu_0 - \mu_1)^2$ [which goes up to $\approx 0.1 d^2$ for our choice of levels], defining $\lambda_{z} = 1 - 3 \cos^2 \Theta_0$ (tunable via $\Theta_0$ between $-2$ and $1$), and
$\lambda_{xy} = - \frac{2 (b+1)}{3} - \frac{4 b +1}{3} \lambda_{z}$ [see shaded area in Fig.~\ref{fig:SPT}(b)], we obtain Eq.~(\ref{eq:XXZ}) with $J_z(\Phi) = 1 -  (1 - \lambda_{z}) \sin^2 \Phi$ and $J_{xy}(\Phi) = 1- (1-\lambda_{xy}) \sin^2 \Phi$. As desired, at nearest-neighbor level, these expressions for $J_z(\Phi)$ and $J_{xy}(\Phi)$ reproduce  
Fig.\ \ref{fig:SPT}(a).
%
\begin{figure}[b]
\includegraphics[width=0.99 \columnwidth]{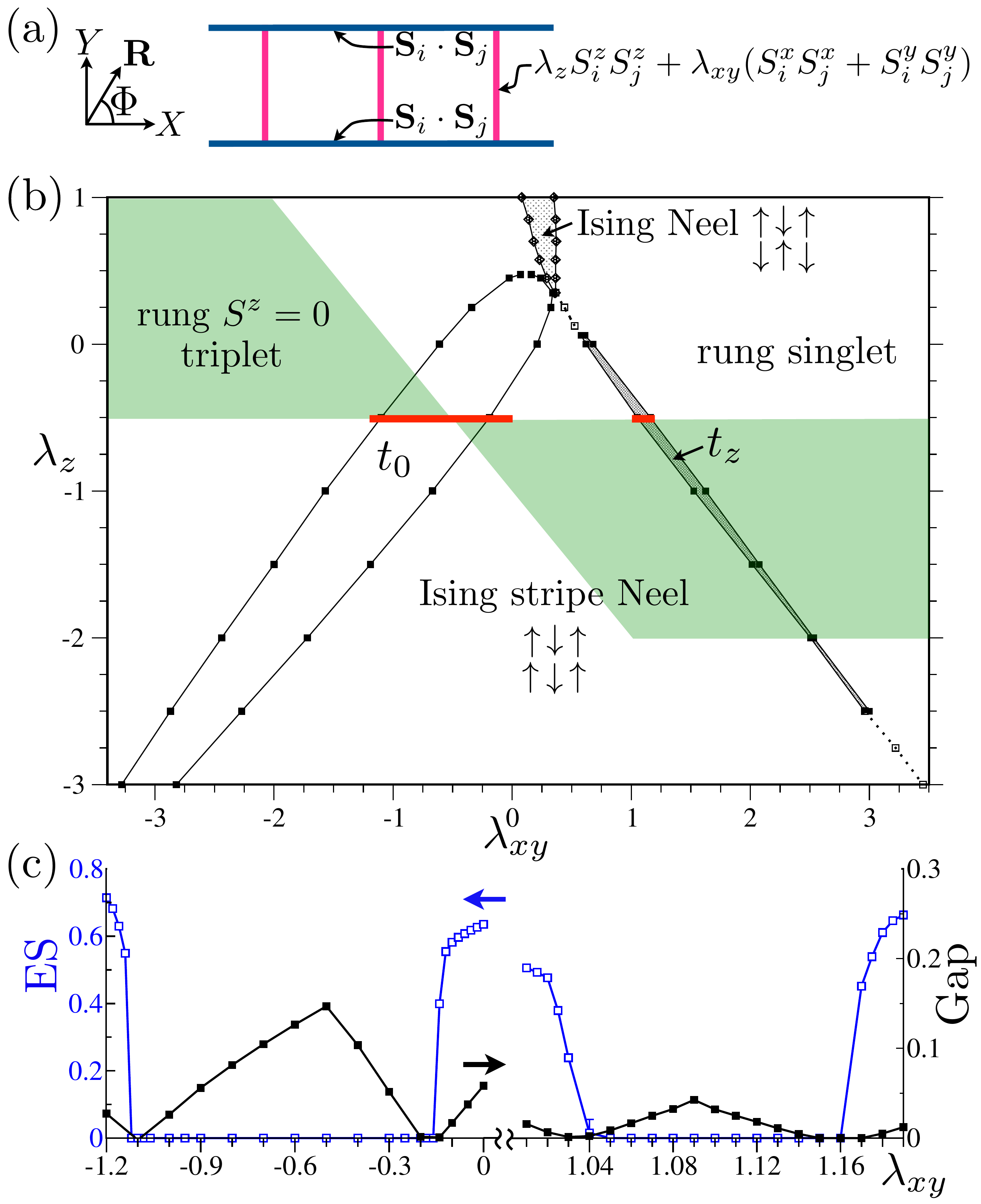}
\caption{(color online).  Spin ladder of Ref.\ \cite{liu12b} and its phase diagram showing SPT phases in the presence of long-range interactions. (a) The nearest-neighbor model. (b)
Phase diagram in the presence of long-range interactions. 
The shaded area indicates points achievable with the simple configuration of Fig.~\ref{fig:levs}(a). The shaded area does not extend past the limits of the vertical axis, while it extends infinitely far along the horizontal axis. (c) Entanglement splitting (open boxes, left axis) and energy gap (solid boxes, right axis) of the SPT phases along  $\lambda_z = -0.5$ cuts (bold, red lines) shown in (b).
}
\label{fig:SPT}
\end{figure}

In our implementation using polar molecules, the nearest-neighbor interactions are replaced by dipolar  interactions, which 
give rise to nontrivial  longer range corrections. In order to investigate  the role of these corrections, we numerically calculate the  phase diagram of the spin ladder with long-range interactions, as shown in Fig.~\ref{fig:SPT}(b), using DMRG on 200 rungs with smooth boundary conditions \cite{vekic93}.
 By performing a finite-size scaling using systems with up to 400 rungs, we estimate the finite-size effects to be 
 comparable to the size of the symbols in Fig.~\ref{fig:SPT}(b). The phase diagram is qualitatively similar to the nearest-neighbor case \cite{liu12b} and 
 exhibits the same six phases, including 
 the two SPT phases.  In Ref.~\cite{liu12b}'s language, the four nontopological phases are the Ising Neel, the Ising stripe Neel, and two product phases of rung singlets and $S^z = 0$ rung triplets.  The remaining two phases, $t_0$ and $t_z$, are two out of seven nontrivial SPT phases protected by $D_2 \times \sigma$ \cite{liu12b}. The $t_0$ phase can be connected to the Haldane phase \cite{haldane83} and to the AKLT state \cite{affleck87} by treating the triplet states on each rung as a spin-1 particle. 
 Meanwhile, in the nearest-neighbor case, $t_z$ is obtained from $t_0$ by taking $\ket{\u} \rightarrow - \ket{\u}$ on one of the legs, which is the $\lambda_{xy} \rightarrow -\lambda_{xy}$ symmetry of the nearest-neighbor phase diagram \cite{liu12b}.
Long-range interactions break this symmetry  and, in particular, reduce the size of the $t_z$ phase relative to the $t_0$ phase as a result of substantial next-nearest-neighbor $S^x_i S^x_j + S^y_i S^y_j$ interactions for $\lambda_{xy} > 0$.

We observe that the $t_z$ phase is sensitive to artificial cutoffs in the interaction range, so we use matrix product operators within DMRG 
to provide an efficient description of interactions \textit{without a cutoff}; instead, we fit the
long-range interactions to a sum of exponentials \cite{mcculloch08,crosswhite08,pirvu10,stoudenmire10,supp}.
The boundaries of the Ising Neel and Ising stripe Neel phases were obtained by calculating the corresponding order parameters. We identified the two rung phases by calculating $\langle S^x_i S^x_j + S^y_i S^y_j \rangle$ on the rungs and by verifying that the gap doesn't close as $|\lambda_{xy}|$ increases to large values where the system is ultimately exactly solvable.
The boundaries of the SPT phases were obtained by computing the entanglement splitting, which we define as
$\textrm{ES} = \sum_{j = \textrm{odd}} \left( w_j - w_{j+1} \right)$,
where $w_j$ are the eigenvalues of the reduced density matrix for a bipartition at the center of the 
system, sorted from largest to smallest. Due to their two-fold degenerate entanglement spectrum \cite{pollmann10}, SPT phases have $\textrm{ES} = 0$, as shown in Fig.~\ref{fig:SPT}(c). 
We have also verified that all phases are gapped.  Interestingly, the energy gap in the SPT phases, shown in Fig.~\ref{fig:SPT}(c), exhibits a cusp indicative of a level crossing,
which deserves further investigation. Finally, using again systems with 200 rungs, we added to the Hamiltonian a small term 
$\propto S^x_i \pm S^x_j$, where $i$ and $j$ are sites on 
the same edge rung.  Referring to operators that split the edge degeneracy as active operators, we verified the prediction \cite{liu12b} that $S^x_i + S^x_j$ is an active operator for the $t_0$  phase while $S^x_i - S^x_j$ is not, and vice versa for the $t_z$ phase. 


\textit{Experimental considerations.}---
As suggested in Ref.~\cite{liu12b}, an SPT phase can be classified by finding its active operators, i.e.~those operators that split the edge degeneracy.  
We propose to diagnose this splitting by measuring an active operator in linear response to the application of that same operator at frequency $\omega$ and looking for the zero-bias ($\omega = 0$) peak. 
Repeating the same procedure for inactive operators will yield no zero-bias peak. 
In our implementation,  a $z$ magnetic field  proportional to $\mu^2_0 - \mu^2_1$ naturally arises at the edges from dipole-dipole interactions \cite{supp}. Since such a field constitutes an active operator of the $t_0$ and $t_z$ phases, it is natural to probe the response of the system by tuning $\mu^2_0 - \mu^2_1$ with the DC electric field.
In combination with a spectroscopic verification of the bulk gap, the response to active operators allows one to detect and classify SPT phases. A more 
modest 
first experimental step could be to use a Ramsey-type experiment \cite{hazzard12b} to benchmark how accurately the molecules emulate
the desired Hamiltonians. 

Polar alkali dimers have hyperfine structure $H_\textrm{hf}$ \cite{aldegunde08,gorshkov11c}, which we have ignored so far. We will illustrate how to deal with $H_\textrm{hf}$ for the specific  case of 
$S=1/2$. Assuming that microwave Rabi frequencies $\Omega_i$ are much larger than $H_\textrm{hf}$, 
we can project the hyperfine structure on dressed states $\ket{\u}$ and $\ket{\d}$. The necessary conditions $H_\textrm{hf} \ll \Omega_i \ll B$ are easy to satisfy: for example, in ${}^{40}$K${}^{87}$Rb, $H_\textrm{hf} \sim (2 \pi) 1$ MHz and the rotational constant is $B \sim (2 \pi) 1$ GHz. The simplest situation arises when 
an applied magnetic field -- of a strength already experimentally used \cite{ni08} -- makes 
$\bra{\u}H_\textrm{hf}\ket{\u}$ and $\bra{\d}H_\textrm{hf}\ket{\d}$ diagonal in the same basis of decoupled nuclear spins. The nuclear-spin degree of freedom can then be eliminated by working with a single state from this basis.
For smaller magnetic fields, one could prepare the system in any pair of non-orthogonal eigenstates of $\bra{\u}H_\textrm{hf}\ket{\u}$ and $\bra{\d}H_\textrm{hf}\ket{\d}$. 
An imperfect overlap of these 
two states 
will effectively reduce the transition dipole moment between $\ket{\u}$ and $\ket{\d}$, resulting in 
an additional control knob of the interactions.





Controlling tens of
independent microwave frequencies in the frequency range required by
our proposal is straightforward \cite{dian08}.
The two uncertainties involved are in the generation of the microwaves and in the coupling to molecules. 
The latter is dominant: current
ultracold molecule 
experiments observe only 0.1$\%$ deviations in
their $\sim 1$ms microwave pulses without any particular optimization 
\cite{Note4},
 and this is expected to be independent of the number
of microwaves applied. Polarization control is more challenging.
However, this should be attainable, for example, by simply interfering
the outputs of two independently controlled microwave horns.

\textit{Outlook.}---
While dipolar interactions did not destroy the SPT phases in our example, quantum magnets with long-range interactions have recently been shown to harbor novel, unusual, and often dimension-specific physics \cite{deng05, hauke10,peter12,koffel12,wall12,nebendahl12,cadarso12}. The polar-molecule experiment we propose could therefore help guide the theoretical understanding of these effects in 2D and 3D systems -- including SPT phases -- where efficient numerical methods are not available. In fact, the classification of SPT phases is yet to be extended to models with long-range interactions.
Finally, we expect  our methods to be immediately extendable to other dipole-dipole interacting systems such as Rydberg atoms \cite{saffman10,schauss12}, magnetic atoms \cite{aikawa12,lu12}, and magnetic defects in solids \cite{childress06,balasubramanian09,weber10}.

We thank J.\ Preskill, J.\ Ye, D.\ Jin, M.\ Lukin, N.\ Yao, S.\ Michalakis,   A.\ Turner, N.\ Schuch, N.\ Lindner, G.\ Evenbly, M.\ Baranov, J.\ Taylor, S.\ Stellmer, W.\ Campbell, M.\ Foss-Feig, M.\ Hermele, V.\ Gurarie, X.-G.\ Wen,  Z.-X.\ Liu, and M.\ Oshikawa for discussions. This work was supported by NSF, IQIM, NRC, AFOSR, ARO, 
the ARO-DARPA-OLE program, and  the Lee A. DuBridge and Gordon and Betty Moore foundations.  
SRM and KRAH thank KITP for hospitality. We acknowledge the use of the Janus supercomputer facilities at CU Boulder. 
This manuscript is the contribution of NIST and is not subject to U.S. copyright. 






\begin{thebibliography}{999}
\bibitem[S1]{XNote1}%
  \BibitemOpen
  \bibinfo {note} {We will either be working at a constant total $S^z$ or will
  drive the transition between $\mathinner {|{\delimiter "3222378 }\delimiter
  "526930B }$ and $\mathinner {|{\delimiter "3223379 }\delimiter "526930B }$
  with an additional field.}\BibitemShut {Stop}%
\end{thebibliography}

%

\clearpage

\title{Supplementary online material to the manuscript:\\``Topological phases in ultracold polar-molecule quantum magnets"}

\author{Salvatore R. Manmana}
\affiliation{Institute for Theoretical Physics, University of G\"ottingen, D-37077 G\"ottingen, Germany}
\affiliation{JILA, NIST and Department of Physics, University of Colorado, Boulder, CO 80309, USA}
\affiliation{Kavli Institute for Theoretical Physics, University of California, Santa Barbara, CA 93106, USA}
\author{E. M. Stoudenmire}
\affiliation{Department of Physics and Astronomy, University of California, Irvine, CA 92697, USA}
\author{Kaden R. A. Hazzard}
\affiliation{JILA, NIST and Department of Physics, University of Colorado, Boulder, CO 80309, USA}
\affiliation{Kavli Institute for Theoretical Physics, University of California, Santa Barbara, CA 93106, USA}
\author{Ana Maria Rey}
\affiliation{JILA, NIST and Department of Physics, University of Colorado, Boulder, CO 80309, USA}
\author{Alexey V. Gorshkov}
\affiliation{Institute for Quantum Information \& Matter, California Institute of Technology, Pasadena, CA 91125, USA}

\maketitle

\section{Details on the molecular physics}

\subsection{Derivation of Eq.~(2) 
in the main text}

Defining $\left\{|0\rangle,|1\rangle,|2\rangle\right\} = \left\{|0,0\rangle, |1,-1\rangle, |1,1\rangle\right\}$
and $n_m = |m\rangle \langle m|$, the dipole-dipole interaction between molecules $i$ and $j$ that are $\mathbf{R} = (R,\theta,\phi)$ apart is
\ba
R^3 H_{ij}   
          &=& (1- 3 \cos^2 \theta) \label{Xeq:3lev}  \\
        &&\times  \Big[(\mu_0 n_0 + \mu_1 n_{1}+ \mu_1 n_2) (\mu_0 n_0 + \mu_1 n_{1}+ \mu_1 n_2) \nonumber \\
           &&  - \tfrac{1}{2} \mu_{01}^2 (|01\rangle \langle10| + |02\rangle \langle 2 0| + \textrm{h.c.})\Big]  \nonumber \\
        && +  \mu_{01}^2 \tfrac{3}{2} \sin^2 \theta \left[e^{i 2 \phi}  (|01\rangle \langle 2 0| + |10\rangle \langle 0 2|) +\textrm{h.c.}\right]. \nonumber
\ea
Projecting on 
$\ket{\u} = \ket{0}$ and $\ket{\d} = \alpha \ket{-1} - \beta \ket{1}$, 
keeping only the terms that conserve the total $S^z$, 
and dropping a constant, we obtain Eq.~(2) 
in the main text with an additional term $(1 - 3 \cos^2 (\Phi-\Phi_0) \sin^2 \Theta_0) (\mu_0^2 - \mu_1^2) (S^z_i + S^z_j)/3$ on the right-hand side.
For a lattice 
with a single-site or two-site (as in the ladder) unit cell, neglecting edge effects, the additional term is a uniform magnetic field, which is irrelevant since $\sum_i S^z_i$ is conserved~\cite{XNote1}. 
On the other hand, if one is interested in studying edges in the presence of nonzero $J_z$, 
one can choose a level configuration where $\mu_0 = - \mu_1 \neq 0$, so that the additional term vanishes. This is achieved, for example, for $\{\ket{\u},\ket{\d}\} = \{|1,0\rangle,\alpha |2,-1\rangle - \beta |2,1\rangle\}$ at $dE/B = 5.072$ or for $\{\ket{\u},\ket{\d}\} = \{|2,0\rangle,\alpha |2,-1\rangle - \beta |2,1\rangle\}$ at $dE/B = 10.535$. Since a $z$ magnetic field is an active operator for both the $t_0$ and the $t_z$ phases  \cite{Xliu12b}, tuning $E$ away from the value where $\mu_0 = - \mu_1$ allows one to controllably turn on active operators and, thus, probe the nature of the edge states, as discussed in the main text.

\subsection{Details behind Fig.~2(b) 
 in the main text}

We number the states in Fig.~2(b) 
 in the main text in the  left-to-right and  bottom-to-top order as
$\left\{|1\rangle, |2\rangle,|3\rangle, \dots, |15\rangle\right\} = \left\{|0,0\rangle, |1,0\rangle, |1,1\rangle,\dots, |4,4\rangle\right\}$. We then write the three dressed states as $|p\rangle = \sum_{j(p)} \sqrt{x_{j}} |j\rangle$, where $x_j > 0$, $\sum_{j(p)} x_j = 1$, $p = 0, \pm 1$, and $j(p)$ means that $j$ runs only over the 5 states belonging to $|p\rangle$ in Fig.~2(b). 
 Whether $|1\rangle$ refers to the bare state or to the dressed state will be clear from the context. Four microwave fields coupling five bare states that make up each dressed state allow one to arbitrarily tune the composition $x_j$ and energy of the dressed state.  Thus 12 microwaves are needed to fully control all 15 $x_j$. Keeping only resonant terms [as in Eq.~(\ref{Xeq:3lev})] and projecting onto dressed states, the dipole-dipole interaction between two molecules that are $\mathbf{R} = (R,\theta,\phi)$ apart is
\ba
&& H_{ij}
=  \frac{1- 3 \cos^2 \theta}{R^3} \Bigg[ \sum_{p,q} A_p A_q |p q\rangle \langle p q| + \sum_p B_p |p p\rangle \langle p p|
\nonumber
\\
&&
+ \sum_{p < q} \frac{J_{p,q}}{2} (|q p\rangle \langle p q| + \textrm{h.c.}) + J_+  (|00\rangle \langle-11| + \textrm{h.c.})
\Bigg],
\label{Xeq:ABJ}
\ea
where $A_p = \sum_{j(p)} x_{j} \mu_{j}$, $B_p = \sum_{i(p) < j(p)} x_{i} x_{j} d_{ij}$, $J_{p,q} = \sum_{i(p),j(q)} x_{i} x_{j} d_{ij}$, $J_{+} = x_2 \sqrt{x_1 x_4} \mu_{12} \mu_{24}$,
\ba
d_{ij} = \left\{\begin{array}{cl} 2 \mu_{ij}^2 &\textrm{ if } M(i) = M(j) \\
- \mu_{ij}^2 & \textrm{ if } |M(i)-M(j)| = 1 \\
0 & \textrm{ otherwise} \end{array} \right.,
\ea
$\mu_{ij} = \langle i|d^{M(i)-M(j)} |j\rangle$, and $\mu_{i} = \langle i|d^0|i\rangle$. 
To allow for $J_{+}$ to be negative, we can simply change the sign of $\sqrt{x_{4}}$ in the expansion of dressed state $|1\rangle$.

To achieve
\ba
\!\!\!\!\!\!\!\! R^3 H_{ij}\! &=&\!  (1- 3 \cos^2 \theta) \!\left[a_1 + a_2 \mathbf{S}_i \cdot \mathbf{S}_j + a_3 (\mathbf{S}_i \cdot \mathbf{S}_j)^2\right], 
\label{Xeq:blbq2}
\ea
we find the constraints on $A_p$, $B_p$, $J_{p,q}$, and $J_+$ in terms of $a_i$ by matching the matrix elements in Eqs.~(\ref{Xeq:ABJ}) and (\ref{Xeq:blbq2}).
It is straightforward to numerically check that, by tuning $x_j$, these constraints 
can be satisfied for arbitrary $\gamma = \textrm{arg}(a_2 + i a_3)$. As an example, setting $x_j = 0$ for $j = 11, \dots, 15$, we are left with just 8 microwaves (two microwaves are needed to couple $|2,1\rangle$ to $|3,3\rangle$), and they give us a solution with 
$\{a_2,a_3\} = \{0.0027, 0.0055\}$
(i.e.\ $\gamma = 1.1$) for $\{x_1,x_2,\dots, x_{10}\} = \{0.0189, 0.1575, 0.2342, 0.5477, 0.9654, 0.7132, 0.0209,$ $0.1293, 0.1972, 0.0157\}$.

It is worth pointing out that, in many geometries, $S^z_i + S^z_j$ and $(S^z_i)^2 + (S^z_j)^2$  in Eq.~(\ref{Xeq:blbq2}) just lead to uniform single-particle energy shifts in the bulk. Such shifts can be easily compensated by tuning the energies of the dressed states. Thus, in cases where one is not worried about introducing $S^z_i$ and $(S^z_i)^2$ terms near the boundary, the conditions used above can be relaxed even further. For example, we can then set $x_j = 0$ for $j = 9, \dots, 15$, and we find a solution -- requiring only 5 microwaves -- with $\{a_2,a_3\} = \{0.0344,0.0041\}$ (i.e.\ $\gamma = 0.12$).

\section{Details on the numerics} 





All density matrix renormalization group (DMRG) calculations were performed on finite ladders with open boundary
conditions in both the $X$ and $Y$ directions. Our calculations conserved total $S^z$, 
and ground states were found by targeting the $S^z=0$ sector. 
In order to attain discarded weights less than $10^{-8}$, we typically kept between 100 and 1000 density-matrix eigenstates. 

To work with long-range interactions in DMRG,  we used the ITensor library \cite{XNote2} 
and expressed our Hamiltonians as matrix-product operators, which can exactly
encode long-range exponentially decaying interactions. We then approximated our dipolar interactions, which decay as $1/R^3$,
as a sum of exponentials. (For technical details of this procedure, see below.) 
Typically, using just 5 exponentials on systems of
up to 400 rungs was sufficient to obtain fits deviating from the exact interactions by less than $10^{-5}$ at any fixed range. 
We also checked convergence of observables: 
for example, the entanglement splitting of 400-rung systems fit with
5 exponentials differed from results with 12 exponentials by less than $10^{-4}$.

The data in Fig.~3(c) 
 of the main text was obtained via finite-system calculations on system sizes typically up to 400 rungs, 
extrapolated to the thermodynamic limit using quadratic fits. All extrapolations were of good quality (errors order of symbol sizes)
except for the entanglement splitting at $\lambda_{xy}=1.04$, which extrapolated to a negative value. The reported value is 
half of the value of the largest system.
To calculate energy gaps for Fig.~3(c), 
we computed the lowest-lying $S^z = 0$ and $S^z = 1$ states,  sequentially 
orthogonalizing against lower lying states within DMRG \cite{XStoudenmire:2012a}. 

The data in Fig.~3(b) 
was primarily obtained on fixed-size systems of 200 rungs. However, to reduce finite-size
effects, we employed smooth boundary conditions \cite{Xvekic93}, smoothly reducing all spin-spin interactions from full strength to zero over a region
of 
$\approx 20$ rungs at each edge. In this way we were able to obtain results similar to open-boundary systems of roughly twice the size.

To verify that we correctly identified the 
rung phases,  we calculated the correlation function $C_{xy} = \langle S^x_i S^x_j + S^y_i S^y_j \rangle$ on the rungs. As expected, we found that $C_{xy} < 0$ $(C_{xy} > 0)$ in the rung singlet (triplet) phase, and $|C_{xy}| \rightarrow 1/2$ in both phases as $|\lambda_{xy}| \rightarrow \infty$  at a fixed $\lambda_z$.

The Ising Neel and 
Ising stripe Neel phases with well-defined order parameters were obtained without 
introducing a symmetry-breaking field. Instead, the choice of the initial state for the DMRG algorithm picked out one of the two ground states, while the other one could be obtained by flipping all the spins in the initial state.

\subsection{Exponentially Decaying Long-Range Interactions with Matrix Product Operators}

In conventional implementations of the DMRG algorithm, the Hamiltonian is treated
as a sum of individual terms with each term projected separately into the local basis used for each DMRG step \cite{Xnoack05}.
If the Hamiltonian contains long-range interactions, this approach requires including each pairwise interaction term such that DMRG
no longer scales linearly in system size even in one dimensional gapped phases.


DMRG has been understood to be a method for optimizing variational wavefunctions known as matrix
product states (MPS) \cite{Xostlund95,Xschollwoeck11}.
The MPS form of the wavefunction suggests that operators may be written in a similar form, known as a matrix product operator (MPO):
\be
\hat{W} = \sum_{\{s,s^\prime\}, \{\alpha\}} W^{s_1 s^\prime_1}_{\alpha_1} W^{s_2 s^\prime_2}_{\alpha_1 \alpha_2} \cdots \ket{s_1 s_2 \cdots} \bra{s^\prime_1 s^\prime_2 \cdots} \ .
\ee
Besides being more convenient to use in MPS-based algorithms, rewriting the Hamiltonian as an MPO offers key technical advantages.
Most remarkably, a finite-bond-dimension MPO can exactly represent a Hamiltonian with exponentially decaying
long-range interactions \cite{Xmcculloch08,Xcrosswhite08,Xpirvu10}.

As a concrete example, the following MPO consisting of only $3\times 3$ matrices
\be
W_j = \begin{bmatrix}
I_j & 0 & 0  \\
S^z_j & \lambda I_j & 0 \\
-h\,S^x_j & J\, S^z_j & I_j
\end{bmatrix} \:
\label{Xeq:lr_isingmpo}
\ee
encodes the exponentially long-range interacting Ising model:
\be
H = J \sum_{i<j}^N S^z_i\  \lambda^{(j-i-1)}\  S^z_j - h\, \sum_j S^x_j \ .
\ee
(Setting $\lambda=0$ restores the conventional nearest-neighbor model.)
Here we notate MPO tensors as matrices of operators.
For example, Eq.~(\ref{Xeq:lr_isingmpo}) indicates that \mbox{$(W_{1,1})^{s_j s^\prime_j} = (I_j)^{s_j s^\prime_j}$}
and  \mbox{$(W_{2,1})^{s_j s^\prime_j} = (S^z_j)^{s_j s^\prime_j}$}.
We also assume open boundary conditions, taking only the last row of Eq.~(\ref{Xeq:lr_isingmpo}) on site 1 and the first column of Eq.~(\ref{Xeq:lr_isingmpo}) on site $N$.
Because the MPO has a finite, system-size-independent bond dimension,  it can be used to
study exponentially decaying long-range interactions within DMRG while retaining linear scaling in system size (in 1D gapped phases).

\subsection{Power-Law Decaying Long-Range Interactions with Matrix Product Operators}

An MPO of finite bond dimension cannot exactly represent a Hamiltonian with power-law
decaying long-range interactions. However, for a large enough exponent $\gamma$, power-law interactions
can be well approximated on a finite-size system as a sum of $N_\text{exp}$ exponentials \cite{Xcrosswhite08,Xpirvu10}:
\be
\frac{1}{|j-i|^\gamma} \simeq \sum_{n=1}^{N_\text{exp}} \chi_n \lambda_n^{|j-i|}  \ , \label{Xeq:fit}
\ee
with $N_\text{exp}$ fairly small.
We use the particularly elegant fitting procedure described in Ref.~\cite{Xpirvu10}.

Again using the Ising model as an example, the Hamiltonian
\be
H = J \sum_{i<j}^N {|i-j|^{-\gamma}}\ S^z_i\, S^z_j   - h\, \sum_j S^x_j
\ee
can be represented within the approximation (\ref{Xeq:fit}) by the following MPO
\be
W_j = \begin{bmatrix}
I_j & 0 & 0 & \cdots & 0  \\
\lambda_1 S^z_j & \lambda_1 I_j & 0 & \cdots & 0 \\
\lambda_2 S^z_j & 0 & \lambda_2 I_j & \cdots & 0 \\
\vdots & \vdots & \vdots & \ddots & \vdots \\
-h\,S^x_j & J \chi_1 \, S^z_j & J \chi_2 \, S^z_j  & \cdots& I_j
\end{bmatrix} \:.
\ee
More compactly, this can be written in block-matrix form as
\be
W_j = \begin{bmatrix}
I_j & 0 & 0   \\
\vec{\lambda} S^z_j & (\mathbf{I} \vec{\lambda}) I_j & 0 \\
-h\,S^x_j & J \vec{\chi} S^z_j  & I_j
\end{bmatrix} \:,
\ee
noting the resemblance of the above to Eq.~(\ref{Xeq:lr_isingmpo}).

\subsection{Power-Law Decaying Long-Range Interactions for Ladder Systems}

The standard approach for applying DMRG to ladder systems is
to map the sites into one dimension through the following ordering:
\begin{center}
\begin{picture}(100,35)
\put(0,0){\makebox(50,32){
\includegraphics[width=140px]{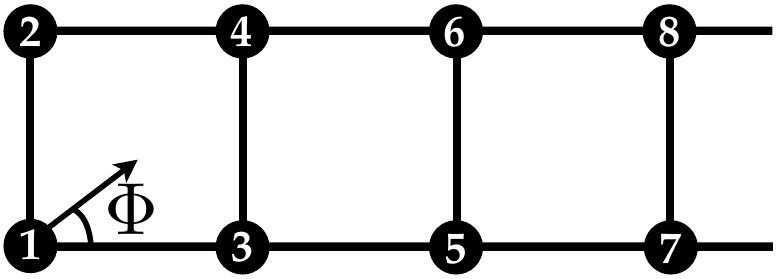}
}}
\end{picture}
\end{center}
We may then study long-range interacting ladder systems within DMRG by applying and generalizing the MPO techniques discussed above.

For example, to encode the same long-range interactions along each leg, but not between the legs, we allow the MPO matrices to be different on each leg:
\begin{align}
W^{(1)}_j & = \begin{bmatrix}
I_j & 0 & 0  & 0 \\
S^z_j & (\mathbf{I} ) I_j & 0 & 0 \\
0 & 0 & (\mathbf{I}  \vec{\lambda}) I_j & 0 \\
0 & J \vec{\chi} S^z_j & 0  & I_j
\end{bmatrix} \\
 & \nonumber \\
W^{(2)}_j & = \begin{bmatrix}
I_j & 0 & 0  & 0 \\
0 & (\mathbf{I} \vec{\lambda} ) I_j & 0 & 0 \\
S^z_j & 0 & (\mathbf{I}) I_j & 0 \\
0 & 0 & J \vec{\chi} S^z_j  & I_j
\end{bmatrix}
\end{align}
A different but related pattern gives only inter-leg interactions. We can include both intra- and inter-leg interactions by combining both patterns in a block-diagonal form.

In this work, we consider Eq.~(2) 
in the main text with $J_z(\Phi) = 1 -  (1 - \lambda_{z}) \sin^2 \Phi$ and $J_{xy}(\Phi) = 1- (1-\lambda_{xy}) \sin^2 \Phi$, where $\Phi$ is the polar angle of the vector connecting the two interacting spins, as shown in the figure above.

To fit these interactions to sums of exponentials, we must 
 perform three fits -- (Heisenberg) intra-leg, $zz$ inter-leg, and $xy$ inter-leg interactions -- to yield the three 
 sets of fitting parameters: $(\vec{\chi}, \vec{\lambda})$, $(\vec{\chi}_z, \vec{\lambda}_z)$, and $(\vec{\chi}_{xy}, \vec{\lambda}_{xy})$.
 

With these definitions, the MPO matrices $W^{(1)}$ on leg 1 and $W^{(2)}$ on leg 2 take the form \cite{XNote3}: 
\begin{widetext}
\begin{align}
W^{(1)}_j & = \left[ \begin{array}{cccccccccccccc}
I_j  \\
S^z_j & (\mathbf{I} ) I_j  \\
S^+_j & 0 & (\mathbf{I} ) I_j  \\
S^-_j & 0 & 0 & (\mathbf{I} ) I_j \\
0 & 0 & 0 & 0 & (\mathbf{I}  \vec{\lambda}) I_j \\
0 & 0 & 0 & 0 & 0 &  (\mathbf{I}  \vec{\lambda}) I_j \\
0 & 0 & 0 & 0 & 0 & 0 & (\mathbf{I}  \vec{\lambda}) I_j  \\
0 & 0 & 0 & 0 & 0 & 0 & 0 & (\mathbf{I} \vec{\lambda}_z) I_j \\
\vec{\lambda}_z S^z_j & 0 & 0 & 0 & 0 & 0 & 0 & 0 & (\mathbf{I} ) I_j \\
0 & 0 & 0 & 0 & 0 & 0 & 0 & 0 & 0 & (\mathbf{I}  \vec{\lambda}_{xy}) I_j \\
0 & 0 & 0 & 0 & 0 & 0 & 0 & 0 & 0 & 0 & (\mathbf{I}  \vec{\lambda}_{xy}) I_j \\
\vec{\lambda}_{xy} S^+_j & 0 & 0 & 0 & 0 & 0 & 0 & 0 & 0 & 0 & 0 & (\mathbf{I} ) I_j \\
\vec{\lambda}_{xy} S^-_j & 0 & 0 & 0 & 0 & 0 & 0 & 0 & 0 & 0 & 0 & 0 & (\mathbf{I}) I_j \\
0 & \vec{\chi} S^z_j & \frac{1}{2} \vec{\chi} S^-_j  & \frac{1}{2} \vec{\chi} S^+_j & 0 & 0 & 0 & \vec{\chi}_z S^z_j & 0 & \frac{1}{2} \vec{\chi}_{xy} S^-_j & \frac{1}{2} \vec{\chi}_{xy} S^+_j  & 0 & 0 & I_j
\end{array}\right] 
\end{align}
\begin{align}
W^{(2)}_j & = \left[ \begin{array}{cccccccccccccc}
I_j  \\
0 & (\mathbf{I}  \vec{\lambda} ) I_j  \\
0 & 0 & (\mathbf{I}  \vec{\lambda}) I_j  \\
0 & 0 & 0 & (\mathbf{I}  \vec{\lambda}) I_j \\
S^z_j & 0 & 0 & 0 & (\mathbf{I} ) I_j \\
S^+_j & 0 & 0 & 0 & 0 &  (\mathbf{I} ) I_j \\
S^-_j & 0 & 0 & 0 & 0 & 0 & (\mathbf{I} ) I_j  \\
\vec{\lambda}_z S^z_j & 0 & 0 & 0 & 0 & 0 & 0 & (\mathbf{I} ) I_j \\
0 & 0 & 0 & 0 & 0 & 0 & 0 & 0 & (\mathbf{I} \vec{\lambda}_z) I_j \\
\vec{\lambda}_{xy} S^+_j & 0 & 0 & 0 & 0 & 0 & 0 & 0 & 0 & (\mathbf{I} ) I_j \\
\vec{\lambda}_{xy} S^-_j & 0 & 0 & 0 & 0 & 0 & 0 & 0 & 0 & 0 & (\mathbf{I} ) I_j \\
0 & 0 & 0 & 0 & 0 & 0 & 0 & 0 & 0 & 0 & 0 & (\mathbf{I} \vec{\lambda}_{xy}) I_j \\
0 & 0 & 0 & 0 & 0 & 0 & 0 & 0 & 0 & 0 & 0 & 0 & (\mathbf{I} \vec{\lambda}_{xy}) I_j \\
0 & 0 & 0 & 0 & \vec{\chi} S^z_j & \frac{1}{2} \vec{\chi} S^-_j  & \frac{1}{2} \vec{\chi} S^+_j  & 0 & \vec{\chi}_z \vec{\lambda}_z S^z_j & 0 & 0 & \frac{1}{2} \vec{\chi}_{xy} \vec{\lambda}_{xy} S^-_j & \frac{1}{2} \vec{\chi}_{xy} \vec{\lambda}_{xy} S^+_j  & I_j
\end{array}\right]
\end{align}
\end{widetext}

\end{document}